\documentclass{ws-p8-50x6-00}
\begin{document}

\title{A new resonance in $K^+\Lambda$ electroproduction: the $D_{13}$(1895) 
       and its electromagnetic form factors}

\author{C. Bennhold, H. Haberzettl}

\address{Center for Nuclear Studies, Department of Physics, The George 
         Washington University, Washington, D.C. 20052, USA}

\author{T. Mart}

\address{Jurusan Fisika, FMIPA, Universitas Indonesia, Depok 16424, 
         Indonesia\\ and \\ Center for Nuclear Studies, Department of 
         Physics, The George 
         Washington University, Washington, D.C. 20052, USA}

\maketitle

\abstracts{New {SAPHIR} $p(\gamma,K^+)\Lambda$ total cross 
section data show a
resonance structure at a total c.m. energy around 1900 MeV. We investigate 
this feature with an isobar model and find that the structure can be well 
explained by including a new $D_{13}$ resonance at 1895 MeV. Such a state 
has been predicted by a relativistic quark model at 1960 MeV with significant 
$\gamma N$ and $K \Lambda$ branching ratios. We demonstrate how the 
measurement of single and double polarization observables can be used to 
obtain additional information on this resonance. Using recent $(e,e' K^+)$ 
JLab data from Hall C we extract the electromagnetic form factors of this 
state.}

\section{Introduction}
The physics of nucleon resonance excitation continues to provide 
a major challenge to hadronic physics\,\cite{nstar} due to the 
nonperturbative nature of QCD at these energies. While methods like 
Chiral Perturbation Theory are not amenable to $N^*$ physics, lattice 
QCD has only recently begun to contribute to this field.  Most of the 
theoretical work on the nucleon excitation spectrum has been performed in 
the realm of quark models. Models that contain three constituent valence 
quarks predict a much richer resonance spectrum\,\cite{NRQM,capstick94}
than has been observed in $\pi N\to \pi N$ scattering experiments. 
Quark model studies have suggested that those ``missing'' resonances may 
couple strongly to other channels, such as the $K \Lambda$ and $K \Sigma$ 
channels\,\cite{capstick98} or final states involving vector mesons.

\section{The Elementary Model}
Using new SAPHIR data\,\cite{saphir98} we reinvestigate the
$p(\gamma, K^+)\Lambda$ process employing an isobar model
described in Ref.\,\cite{fxlee99}.  We are especially interested in
a structure around W = 1900 MeV, revealed in the $K^+ \Lambda$ total 
cross section data for the first time. Guided by a recent coupled-channels 
analysis\,\cite{feuster98}, the low-energy resonance part of
this model includes three states that have been found to have 
significant decay widths into the $K^+\Lambda$ channel, 
the $S_{11}$(1650), $P_{11}$(1710), and $P_{13}(1720)$ resonances. 
In order to approximately account for unitarity corrections at tree-level 
we include energy-dependent widths along with partial branching fractions 
in the resonance propagators\,\cite{fxlee99}. The background part includes 
the standard Born terms along with the $K^*$(892) and $K_1$(1270) 
vector meson poles in the $t$-channel. As in Ref.\,\cite{fxlee99}, we 
employ the gauge method of Haberzettl\,\cite{haberzettl98}
to include hadronic form factors.
The fit to the data was significantly improved by
allowing for separate cut-offs for the background and resonant sector.
For the former, the fits produce a soft value around 800 MeV,
leading to a strong suppression of the background terms while 
the resonant cut-off is determined to be 1900 MeV.  

\begin{table}[!t]
\caption{Comparison between the extracted fractional decay widths and
        the result from the quark model \protect\cite{capstick98,capstick92} 
        for the $S_{11}(1650)$, $P_{11}(1710)$, $P_{13}(1720)$, and 
        $D_{13}(1895)$ resonances.}
\begin{center}
\footnotesize
\renewcommand{\arraystretch}{1.6}
\label{table_cc1}
\begin{tabular}{|l|c|c|c|}
\hline
&&\multicolumn{2}{|c|}{~~~~~~~~~~ $\sqrt{\Gamma_{N^*N\gamma}
\Gamma_{N^*K\Lambda}}/\Gamma_{N^*}$ ($10^{-3}$) ~~~~~~~~~~ } \\[0.8ex]
\cline{3-4} 
\raisebox{2.8ex}{~~~~ Resonance ~~~~} & 
\raisebox{2.8ex}{~~ Status ~~}
& ~~~~~~~ Extracted ~~~~~~~ & Quark Model \\ [0.5ex]
\hline
~~~~ $S_{11}(1650)$ & ****& $-4.83\pm 0.05$ & $-4.26\pm 0.98$ \\
~~~~ $P_{11}(1710)$ & ***& $ ~~1.03\pm 0.17$ & $-0.54\pm 0.12$ \\
~~~~ $P_{13}(1720)$ & ****& $~ 1.17\pm 0.04 $ & $-1.29\pm 0.24$\\
~~~~ $D_{13}(1895)$ & $\dagger$ & $~ 2.29^{+0.72}_{-0.20}$ & $-2.72\pm 0.73$\\
[0.5ex]
\hline
\multicolumn{4}{l}{$\dagger$ Ref. \protect\cite{capstick98} obtains
 a mass of 1960 MeV for this state and relates it
  to the $D_{13}(2080)$,}\\
\multicolumn{4}{l}{given by PDG \protect\cite{pdg} with a ** status.}
\end{tabular}
\end{center}
\end{table}

\section{Results from Kaon Photoproduction: a new $D_{13}$ State at 1895 MeV}

\begin{figure}[!t]
\begin{center}
\epsfig{figure=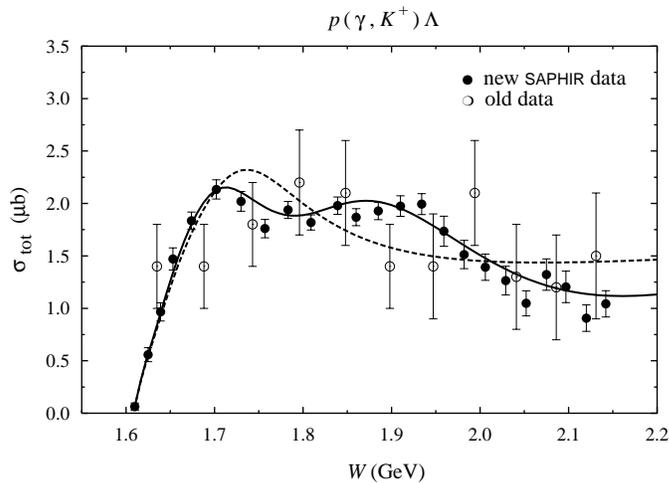,width=90mm}
\vspace{-5mm}
\caption{Total cross section for $K^+\Lambda$ photoproduction
        on the proton. The dashed line shows the model without the 
        $D_{13}(1895)$ resonance, while the solid line is obtained 
        by including the $D_{13}(1895)$ state. The new {\scriptsize SAPHIR} 
        data are from Ref.\,\protect\cite{saphir98}, old data from 
        Ref.\,\protect\cite{old_data}.\label{fig:total}}
\end{center}
\end{figure}

\begin{figure}[!]

\vspace{-25mm}

\epsfig{figure=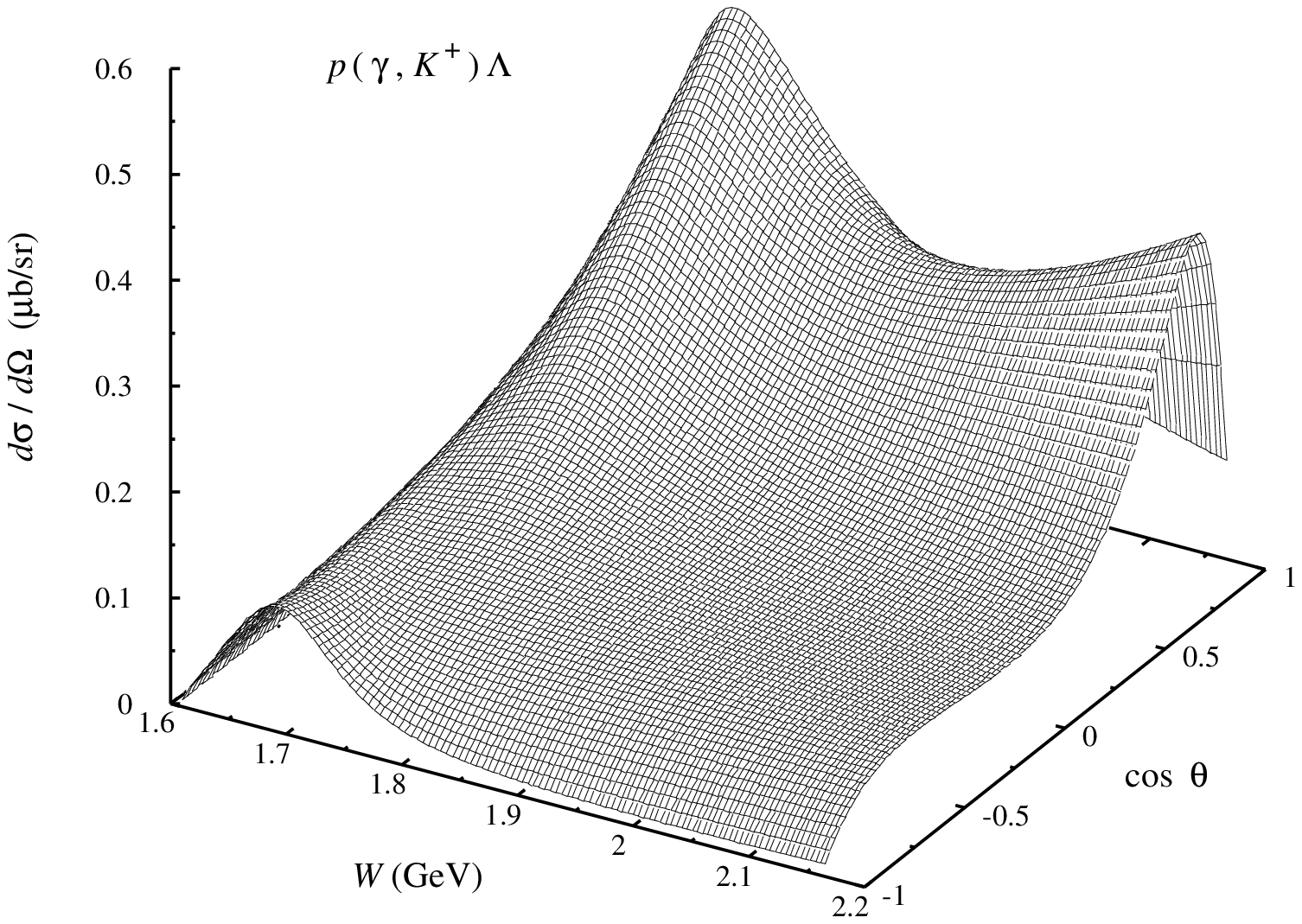,width=120mm}

\vspace{-35mm}

\epsfig{figure=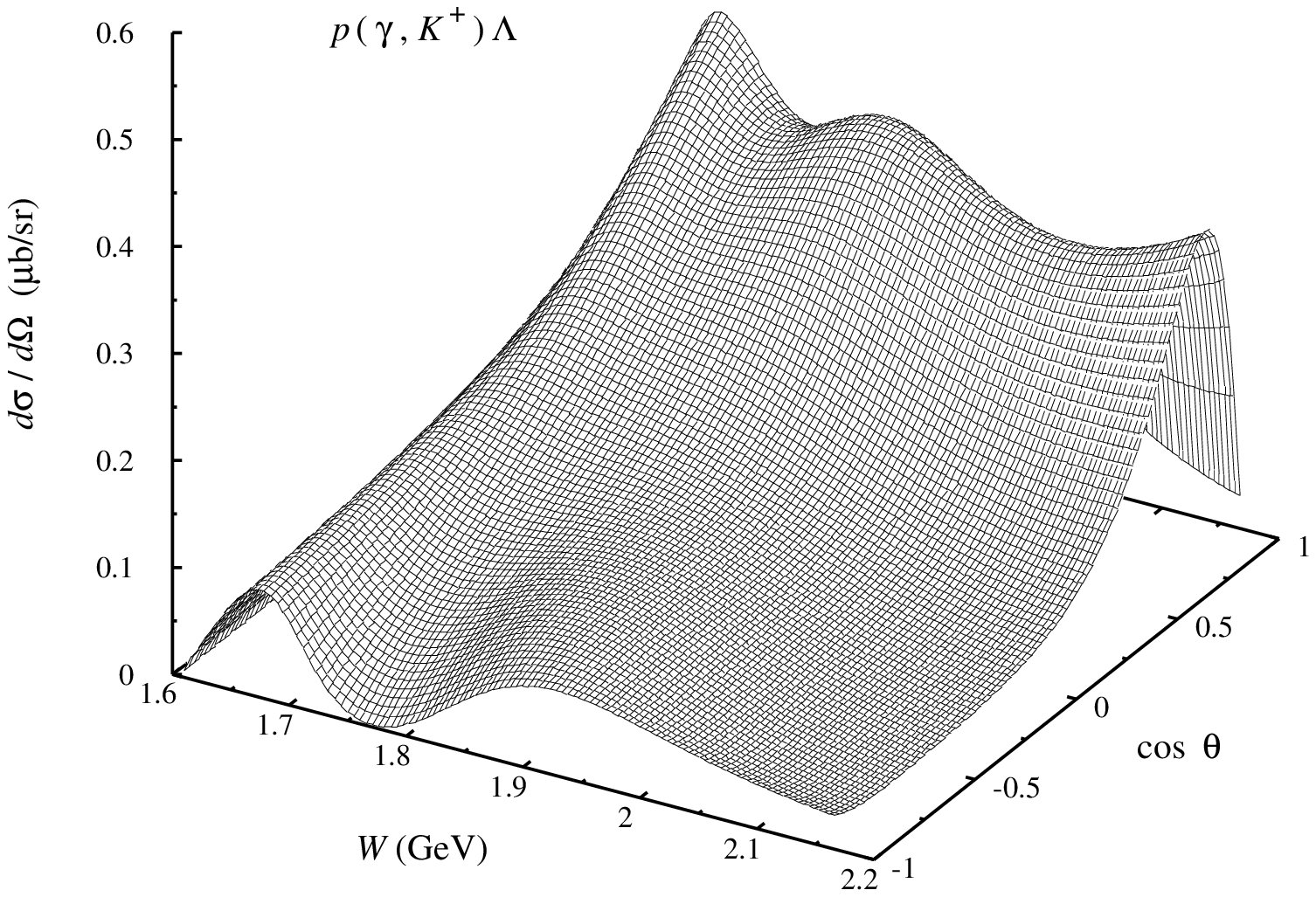,width=120mm}

\vspace{-5mm}

\caption{Differential cross section for the fit without (top) and with
        (bottom) the $D_{13}(1895)$ resonance.\label{fig:dif1}}
\end{figure}

Figure \ref{fig:total} compares our model described above
with the {\small SAPHIR} total cross section data.
Our result shows only one peak
near threshold and cannot reproduce the data at higher energies 
without the inclusion of a new resonance with a mass of around 1900 MeV. 
While there are no 3 - or 4-star isospin 1/2 resonances around 1900 MeV
in the Particle Data Book, several 2-star states are listed,
such as the $P_{13}(1900)$, $F_{17}(1990)$, $F_{15}(2000)$ and 
$D_{13}(2080)$.  On the theoretical side, the constituent quark model by 
Capstick and Roberts\,\cite{capstick94} predicts many new states around 
1900 MeV, however, only few of them have been calculated to have
a significant $K \Lambda$ decay width\,\cite{capstick98}.
These are the $[S_{11}]_3$(1945), $[P_{11}]_5$(1975), $[P_{13}]_4$(1950), 
and $[D_{13}]_3$(1960) states, where the subscript refers to the particular
band that the state is predicted in.  We have performed fits for each of 
these possible states, allowing the fit to determine the mass, width and 
coupling constants of the resonance. While we found that all four states 
can reproduce the structure at $W$ around 1900 MeV, it is only the 
$[D_{13}]_3$(1960) state that is predicted to have a large photocoupling 
along with a sizeable decay width into the $K \Lambda$ channel.
Table \ref{table_cc1} presents the remarkable agreement, up to a sign,
between the quark model predictions and our extracted results for
the $[D_{13}]_3$(1960) state. In our fit, the mass of the $D_{13}$
comes out to be 1895 MeV; we will use this energy to refer
to this state below.

\begin{figure}[!t]
\begin{center}
\epsfig{figure=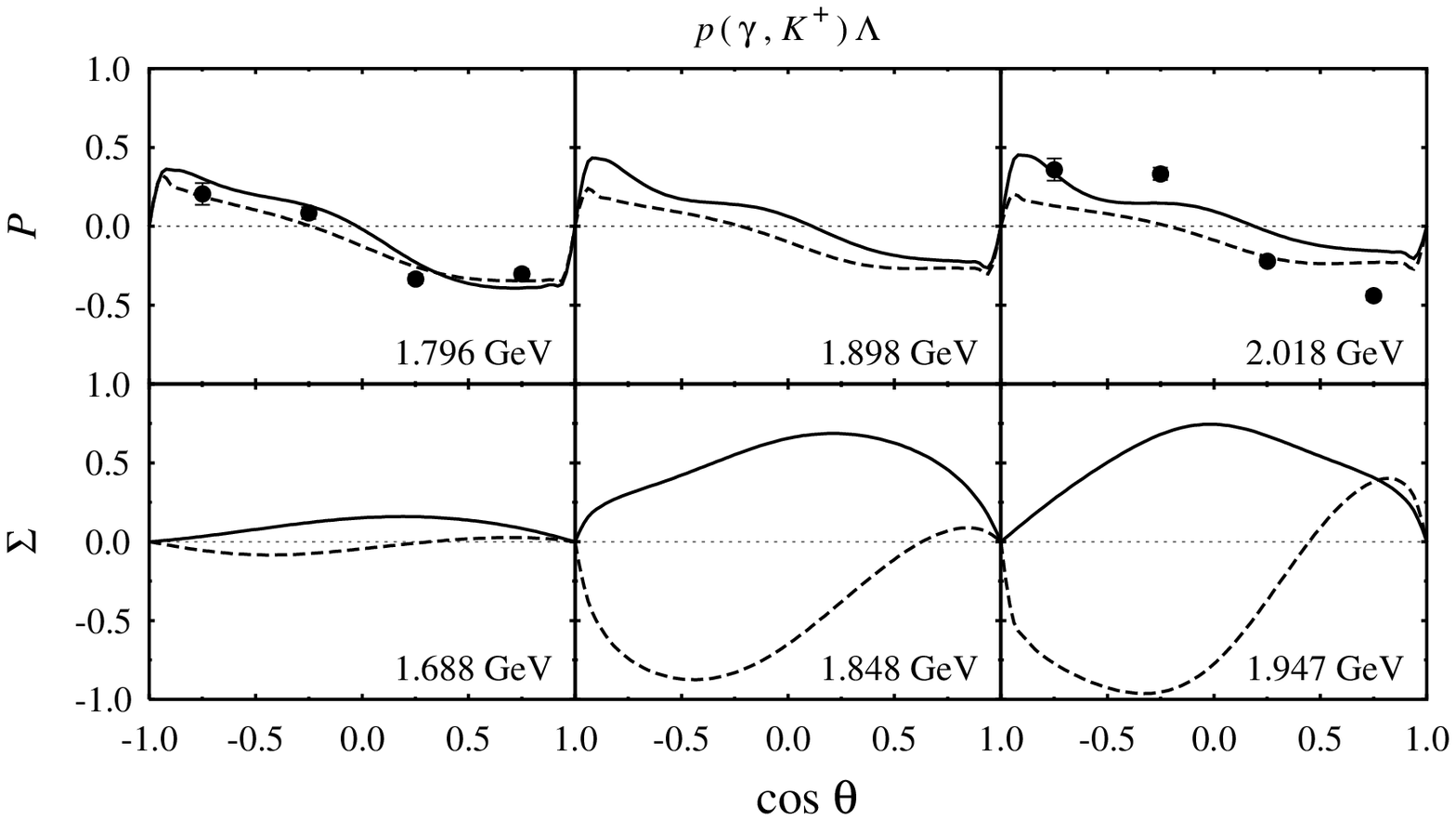,width=95mm}
\caption{Single polarization observables $P_{\Lambda}$ and $\Sigma$ for 
        the reaction $p(\gamma,K^+)\Lambda$.  Notation is the same as in 
        Fig.\,\ref{fig:total}.\label{fig:plamsig}}
\end{center}
\end{figure}

How reliable are the quark model predictions? Clearly, one test is to confront
its predictions with the extracted couplings for the well-established
resonances in the low-energy regime of the $p(\gamma, K^+)\Lambda$ reaction,
the $S_{11}(1650)$, $P_{11}(1710)$ and $P_{13}(1720)$ excitations.
Table \ref{table_cc1} shows that  the magnitudes of the extracted partial 
widths for the $S_{11}(1650)$, $P_{11}(1710)$, and $P_{13}(1720)$ are 
in good agreement with the quark model.  Therefore, even though the amazing 
quantitative agreement for the decay widths of the $D_{13}$ (1895) is 
probably accidental we believe the structure in the SAPHIR data is in all 
likelihood produced by a state with these quantum numbers.  Further evidence 
for this conclusion is found below in our discussion on the recent JLab kaon 
electroproduction data.

As shown in Ref.\,\cite{mart99} the difference between the two 
calculations is much smaller for the differential cross sections. 
Including the $D_{13}$(1960) does not affect the threshold and 
low-energy regime while it does improve the agreement at higher energies.

The difference between the two models can be seen more clearly in 
Fig.\,\ref{fig:dif1}, where the differential cross section is plotted in
a three-dimensional form. As shown by the lower part of Fig.\,\ref{fig:dif1},
the signal for the missing resonance at $W$ around 1900 MeV is most
pronounced in the forward and backward direction. Therefore, in order to
see such an effect in the differential cross section, angular bins should
be more precise for these two kaon directions.

Figure \ref{fig:plamsig} shows that the influence of the new state on the
recoil polarization is rather small for all angles,
which demonstrates that the recoil polarization is not the 
appropriate observable to further study this resonance. On the other hand,
the photon asymmetry of $K^+\Lambda$ photoproduction shows 
larger variations between the two calculations, especially for higher
energies. Here the inclusion of the new state leads to a sign change in 
this observable, a signal that should be easily detectable by experiments
with linearly polarized photons.

\begin{figure}[!t]
\begin{center}
\epsfig{figure=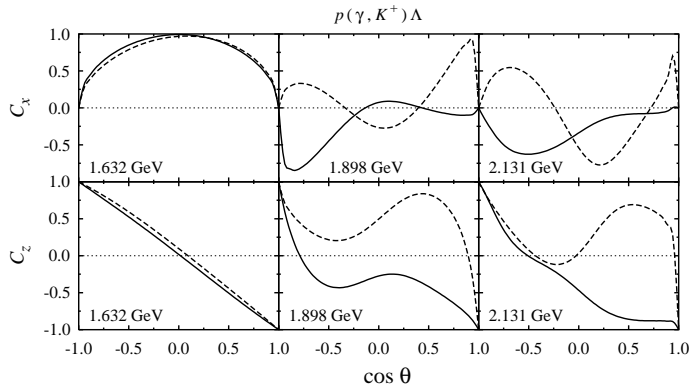,width=95mm}
\caption{Double polarization observables $C_x$ and $C_z$ for the reaction 
        $p(\gamma,K^+)\Lambda$ with a polarized beam and detection 
        of recoil polarization. Notation is the same as in 
        Fig.\,\ref{fig:total}.\label{fig:cxcz}}
\end{center}
\end{figure}

Figure \ref{fig:cxcz} shows double polarization observables
for an experiment with circularly polarized photon and polarized
recoil.  As expected, we find no influence of the $D_{13}$(1895)
at threshold. At resonance energies there are again clear
differences between the two predictions.  

\section{Results from Kaon Electroproduction: 
         Electromagnetic Form Factors of the $D_{13}$(1895)}

All previous descriptions of the kaon electroproduction\,\cite{previous} 
process have performed fits to both photo- and electroproduction data 
simultaneously, in an attempt to provide a better constraint on the 
coupling constants.  This method clearly runs the danger
of obscuring - rather than clarifying - the underlying production mechanism.
For example, anomalous behavior of the response functions in a 
certain $k^2$ range would be parameterized into the effective coupling
constants, rather than be expressed in a particular form factor.
Here, we adopt the philosophy used in pion electroproduction
over the last decade: we demand that the kaon electroproduction
amplitude be fixed at the photon point by the fit to the photoproduction data.
Thus, all hadronic couplings, photocouplings and hadronic form factors are 
fixed, the only remaining
freedom comes from the electromagnetic form factors of the included
mesons and baryons.

Extending our isobar model of Ref.\,\cite{fxlee99} to finite $k^2$ requires 
the introduction of additional contact terms in the Born sector in order to 
properly incorporate gauge invariance\,\cite{haberzettl99}.
We choose standard electromagnetic form factors for the nucleon 
\cite{gari92}, for the hyperons we use the hybrid vector meson dominance 
model\,\cite{williams97}. We use the monopole form factors for the meson
resonances, where their cut-offs are taken as free parameters, determined
to be $\Lambda =1.107$ GeV and $0.525$ GeV for the $K^*$(892) and $K_1$(1270),
respectively.
That leaves the resonance form factors to be determined which in 
principle can be obtained from pion electroproduction. In practice, the 
quality of the data at higher W has not permitted such an extraction. 
For the $S_{11}$(1650) state, we use a parameterization given by 
Ref.\,\cite{penner}. For the $P_{11}$(1710), $P_{13}(1720)$ and 
$D_{13}$(1895) states we adopt the following functional form
for their Dirac and Pauli form factors $F_1$ and $F_2$:
\begin{eqnarray}
  F(k^2) &=& \left( 1-\frac{k^2}{\Lambda^2} \right)^{-n} \, ,
\end{eqnarray}
with the parameters $\Lambda$ and $n$ to be determined by the kaon 
electroproduction data.
The resulting parameters are listed in Table \ref{table_ff}.

\begin{table}[!t]
\caption{Parameters for the $P_{11}(1710)$, $P_{13}(1720)$, and 
        $D_{13}(1895)$ form factors.}
\begin{center}
\footnotesize
\renewcommand{\arraystretch}{1.4}
\label{table_ff}
\begin{tabular}{|l|c|c|c|c|}
\hline
~~~~~~~~Resonance & $\Lambda_1$ (GeV) 
& $n_1$ & $\Lambda_2$ (GeV) & $n_2$ \\[0.8ex]
\hline
~~~~~~~~$P_{11}(1710)$~~~~~~~~ & ~~~~$1.37$~~~~ & ~~~~$4$~~~~ & 
~~~~$-   $~~~~ & ~~~~$-$ \\
~~~~~~~~$P_{13}(1720)$~~~~~~~~ & ~~~~$2.00$~~~~ & ~~~~$1$~~~~ & 
~~~~$2.00$~~~~ & ~~~~$3.31$ \\
~~~~~~~~$D_{13}(1895)$~~~~~~~~ & ~~~~$0.36$~~~~ & ~~~~$4$~~~~ & 
~~~~$1.21$~~~~ & ~~~~$4$ \\    
[0.5ex]
\hline
\end{tabular}
\end{center}
\end{table}

\begin{figure}[!t]
\begin{center}
\epsfig{figure=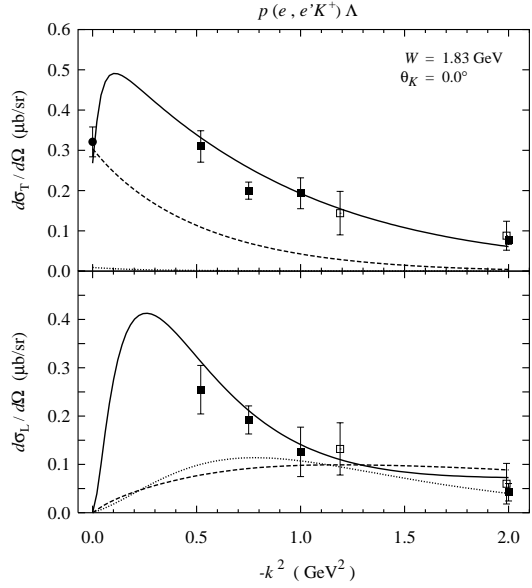,width=75mm}
\caption{Transverse and longitudinal cross section for $p(e,e'K^+)\Lambda$. 
         The dotted line shows the contributions from the background terms,
         otherwise the notation for the curves 
        is the same as in Fig.\,\ref{fig:total}. Solid squares show the
        new JLab data \protect\cite{gabi98}, open squares are the old data 
        \protect\cite{brauel}. In the transverse cross section a 
        photoproduction datum (solid circle) is shown for comparison.
         \label{fig:jlab}}
\end{center}
\end{figure}

Figure \ref{fig:jlab} shows the result of our fit. Clearly, the amplitude
that includes the $D_{13}(1895)$ resonance yields much better agreement with 
the new experimental data\,\cite{gabi98} from Hall C at JLab. The model 
without this resonance produces a transverse cross section which drops
monotonically as a function of $-k^2$, while in the longitudinal case
this model dramatically underpredicts the data for small momentum transfer.
With a $W=1.83$ GeV the data are close in energy to the new state,
thus allowing us to study the $-k^2$ dependence of its form factors.
The contribution of the Born terms is neglegibly small for the
transverse cross section but remains sizeable for the longitudinal one.
We point out that without the $D_{13}(1895)$ we did not find a reasonable
 description of the JLab data, even if
we provided for maximum flexibility in the functional form of the
other resonance form factors.  The same holds true if the new
resonance is assumed to be an $S_{11}$ or a $P_{11}$ state. Even 
including an additional $P_{13}$ state around 1900 MeV
does not improve the fit to the electroproduction data. 
It is only with the interference of two form 
factors given by the coupling structure of a different spin-parity 
state, viz.\,$D_{13}$, that a description becomes possible. 
We therefore find that these new kaon electroproduction data
provide additional evidence supporting our suggestion that the 
quantum numbers of the new state indeed correspond to a $D_{13}$.

\begin{figure}[!t]
\begin{center}
\epsfig{figure=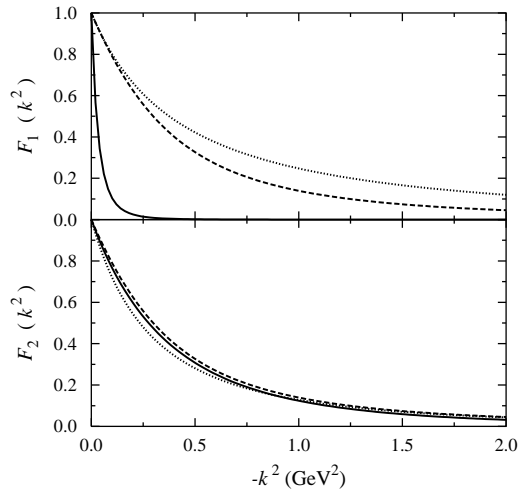,width=70mm}
\caption{Comparison between the form factors of the $D_{13}(1895)$ 
         (solid lines), the $\Delta (1232)$\,\protect\cite{penner} 
         (dashed lines), and the proton\,\protect\cite{gari92}
         (dotted lines).\label{fig:form}}
\end{center}
\end{figure}

The form factors extracted for the $D_{13}(1895)$ are shown in 
Fig.\,\ref{fig:form}, in comparison to the Dirac and Pauli form factors
of the proton and those of the $\Delta$(1232). While the $F_2(k^2)$
form factors look similar for all three baryons, $F_1(k^2)$ of the 
$D_{13}(1895)$ resonance falls off dramatically at small $k^2$. 
It is the behavior of this form factor that leads to the structure 
of the transverse and longitudinal cross sections at a $-k^2$ = 0.2 - 0.3 
GeV$^2$; at higher $k^2$ both response functions are dominated by 
$F_2(k^2)$.  The experimental exploration of the small $k^2$ regime
could therefore provide stringent constraints on the extracted form factors.

\section{Conclusion}

We have investigated a structure around $W= 1900$ MeV
in the new {\small SAPHIR} total cross section data in 
the framework of an isobar model and found that the data can 
be well reproduced by including a new
$D_{13}$ resonance with a mass, width and coupling parameters in good 
agreement with the values predicted by the recent quark model calculation
of Ref.\,\cite{capstick98}.
To further elucidate the role and nature of this state we suggest
measurements of the polarized photon asymmetry around $W = 1900$ MeV
for the $p(\gamma, K^+)\Lambda$ reaction.  Furthermore, we extended
our isobar description to kaon electroproduction by allowing
only electromagnetic resonance  transition form factors to vary.
Employing the new JLab Hall C $p(e, e' K^+)\Lambda$ data at $W=1.83$ GeV
we find that a description of these data is only possible when the new
$D_{13}$ state is included in the model.  The dominance of this state at
these energies allowed us to extract its transition form factors,
one of which was found to to be dramatically different from other resonance
form factors.

\section*{Acknowledgments}
We thank Gregor Penner for providing his parameterization
of the $\Delta (1232)$ form factor.
This work was supported by the US DOE grant DE-FG02-95ER-40907 (CB and HH)
and the University Research for Graduate Education (URGE) grant (TM).


\begin{thebibliography}{99}
\bibitem{nstar} {\it Proceedings of the 4th Workshop on $N^*$ Physics}, 
        Washington, D.C., 1997, edited by H.~Haberzettl, C. Bennhold, and 
        W. J. Briscoe, $\pi N$ {\it Newsletter} {\bf 14}, 1 (1998).
\bibitem{NRQM} N. Isgur and G. Karl, {\it Phys. Lett.} B {\bf 72}, 109 (1977);
        {\it Phys. Rev.} D {\bf 23}, 817 (1981); R.~Koniuk and N. Isgur, 
        {\it Phys. Rev.} D {\bf 21}, 1868 (1980).
\bibitem{capstick94} S. Capstick and W. Roberts, {\it Phys. Rev.} D {\bf 49},
        4570 (1994).
\bibitem{capstick98} S. Capstick and W. Roberts, {\it Phys. Rev.} 
                     D {\bf 58}, 074011 (1998).
\bibitem{saphir98} {\small SAPHIR} Collaboration:
        M.Q. Tran {\it et al}., {\it Phys. Lett.} B {\bf 445}, 20 (1998).
\bibitem{fxlee99} F.X. Lee, T. Mart, C. Bennhold, and L.E. Wright,
                  `Quasifree Kaon Photoproduction on Nuclei',
                  {\sf nucl-th/9907119}.
\bibitem{feuster98} T. Feuster and U. Mosel, {\it Phys. Rev.} C {\bf 58}, 457 
                    (1998); {\it Phys. Rev.} C {\bf 59}, 460 (1999).
\bibitem{haberzettl98} H. Haberzettl, {\it Phys. Rev.} C {\bf 56}, 2041 
                 (1997); H. Haberzettl, C. Bennhold, T. Mart, and T. Feuster, 
                 {\it Phys. Rev.} C {\bf 58}, R40 (1998). 
\bibitem{capstick92} S. Capstick, {\it Phys. Rev.} D {\bf 46}, 2864 (1992).
\bibitem{pdg} Particle Data Group: 
        C. Caso {\it et al}., {\it Eur. Phys. J.} C {\bf 3}, 1 (1998).
\bibitem{old_data} ABBHHM
        Collaboration, {\it Phys. Rev.} {\bf 188}, 2060 (1969).
\bibitem{mart99} T. Mart and C. Bennhold, `Evidence for a missing 
                 nucleon resonance in kaon photoproduction',
                 {\sf nucl-th/9906096}.
\bibitem{previous} R.A. Williams, C.-R. Ji, and S.R. Cotanch,
                   {\it Phys. Rev.} C {\bf 46}, 1617 (1992);
                   T. Mart, C. Bennhold, and C. E. Hyde-Wright, 
                   {\it Phys. Rev.} C {\bf 51}, R1074 (1995);
                   J.C. David, C. Fayard, G.H. Lamot,
                   and B. Saghai, {\it Phys. Rev.} C {\bf 53}, 2613 (1996).
\bibitem{haberzettl99} H. Haberzettl, T. Mart, and C. Bennhold, in
         preparation.
\bibitem{gari92} M.F. Gari and W. Kr\"umpelmann, {\it Phys. Rev.} 
                 D {\bf 45}, 1817 (1992).
\bibitem{williams97} R.A. Williams and T.M. Small, {\it Phys. Rev.} 
                     C {\bf 55}, 882 (1997).
\bibitem{penner} G. Penner, T. Feuster, and U. Mosel, `Pion Electroproduction 
                 and Pion Induced Dileptonproduction on the Nucleon',
                 {\sf nucl-th/9802010}.
\bibitem{gabi98} G. Niculescu {\it et al.}, {\it Phys. Rev. Lett.} {\bf 81},
              1805 (1998).
\bibitem{brauel} P. Brauel {\it et al}., {\it Z. Phys.} C {\bf 3}, 101 (1979).
\end{thebibliography}
\end{document}